\documentclass[12pt, letterpaper, onecolumn, peerreview]{IEEEtran}
\usepackage{graphics}

\usepackage{psfrag}

\usepackage{graphicx,verbatim,fancyhdr}
\usepackage{amssymb}
\usepackage{amsmath}

\newtheorem{theorem}{Theorem}
\newtheorem{lemma}{Lemma}

\title{\large \bf Coding into a source: a direct inverse Rate-Distortion theorem}

\author{Mukul Agarwal, Anant Sahai, and Sanjoy Mitter %
\thanks{Mukul Agarwal is an EECS student at MIT. The core of this work
  was performed largely while he was visiting Prof.~Sahai at Wireless Foundations at UC Berkeley. {\tt\small magar@mit.edu}}
\thanks{Anant Sahai is with Wireless Foundations in EECS at UC Berkeley
        {\tt\small sahai@eecs.berkeley.edu}}%
\thanks{Sanjoy Mitter is with LIDS in EECS at MIT  {\tt\small mitter@mit.edu}}%
}

\begin{document}

\maketitle 

\begin{abstract}
Shannon proved that if we can transmit bits reliably at rates larger
than the rate distortion function $R(D)$, then we can transmit this
source to within a distortion $D$. We answer the
converse question ``If we can transmit a source to within a distortion
$D$, can we transmit bits reliably at rates less than the rate
distortion function?'' in the affirmative. This can be viewed as a
direct converse of the rate distortion theorem.
\end{abstract}


\IEEEpeerreviewmaketitle

\section{Introduction}

In \cite{ShannonLossy}, Shannon proved that if there is a channel with
capacity $C > R(D)$, a source can be transmitted to within a
distortion $D$ reliably over this channel ($R(D)$ is the rate
distortion function for the source) in two steps:
\begin{enumerate}
\item Suppose $C = R(D - \alpha)$. First, source code to within a
  distortion $(D - \frac{\alpha}{2})$ by using random codes.  The
  source code has rate arbitrarily close to $R(D - \frac{\alpha}{2})$.
\item Transmit these bits reliably\footnote{The bounded nature of the
    distortion function only becomes important if we are interested in
    end-to-end expected distortion. If all that is desired is for the
    probability of excess distortion to be arbitrarily small, then no
    such assumptions are needed.} over the channel.
\end{enumerate}
The traditional converse to this separation theorem is proved using
the data-processing inequality and shows that no other joint
source-channel scheme can do any better.

% Let the source be iid. $X_1^\infty$, $X_i \sim p_X$. Denote the
% source code by $\hat{X}_i$ and denote the decoded message at the
% output of the channel by $Y_i$. By doing the above one achieves a
% constraint on the expected distortion:
% \begin{equation}\label{ExpectedDistortion}
% Ed(X_i, Y_i) \leq D \ \forall i
% \end{equation}
% Infact, since source code is a random code, it is easy to see that 
% \begin{equation} \label{BlockDistortion0}
% \frac{1}{n} \sum_{t = 1}^{n} d(X_t, \hat{X}_t) \leq D
% \end{equation}
% with probability which $\to$ 1 as the block length $n \to \infty$.
 
We want to instead ask the converse question at the engineering
level\footnote{Fundamentally, we are asking whether reliable lossless
  communication is necessarily the right primitive that defines
  layering in a multipurpose communication system. Could lossy coding
  serve as an equally good primitive in principle?}: if there is a
``black box'' over which an iid source $X_i \sim p_X$ can be transmitted
to within a distortion level $D$, can we do reliable communication of
bits (in the Shannon sense) over this ``black box'' at rates less than
$R(D)$?

If one assumes that the communication of $X_i$ over the black box
satisfies only an expected distortion constraint $Ed(X_i, \hat{X}_i)
\leq D$, then we \textbf{cannot} guarantee reliable communication. The
black box should be viewed as an attacker and the attacker can do
anything that it wishes as long as it meets the expected distortion
constraint.

Consider an equiprobable binary source $\{0, 1\}$ under the Hamming
distortion.  Suppose the black box is constrained to communicate this
source to within an expected distortion of $0.25$. A possible attacker
could flip a fair coin once at the beginning of time. If it is heads,
then it transmits the symbols perfectly for all time; if it is tails,
it just transmits 0 for all time. It is then easy to see that one
cannot do reliable communication over this attacker at any non-zero
rate, whereas the rate-distortion $R(0.25) > 0$.

Thus, the expected distortion constraint is not sufficient. It turns
out that a block distortion constraint is sufficient. If the attacker
is such that\footnote{For simplicity of notation, the dependence of
  the attacker on block-length $n$ is suppressed. To be precise,
  (\ref{BlockDistortion}) should be interpreted as a family of
  attackers indexed by $n$ such that the probability of excess
  distortion can be made as close to zero as desired by choosing an
  attacker with an appropriately large $n$. This parallels the
  existence result for channel coding.} 
\begin{equation} \label{BlockDistortion}
\Pr \left (\frac{1}{n} \sum_{t = 1}^{n} d(X_t, Y_t) \right ) > D
\to 0 \ \mbox{as } n  \to \infty
\end{equation}
it can be proved that reliable communication is possible over this
attacker at all rates less than $R(D)$. This is the main theorem of
this paper which is stated formally in Section~\ref{MainResultsUnconditional}.

Following \cite{ControlPartI}, one can draw an equivalence between
all rate-distortion problems with a given value of $R(D)$. Consider
the collection of all iid sources and corresponding distortion levels,
$(C_{\beta}, D_{\beta})$ such that $R_{C_{\beta}}(D_{\beta}) = R_0$.
If any one of these sources can be communicated over an attacker such that
the block distortion criterion (\ref{BlockDistortion}) holds, then all
of them can be communicated to within a distortion level $(D_{\beta} +
\delta)$ over this same attacker, for arbitrarily small positive
$\delta$. One way to show this is:
\begin{enumerate}
\item Source code one source to within the distortion level 
$D_{\beta} + \delta$ by using less than $nR_0$ bits.
\item Communicate these $nR_0$ bits reliably by embedding
  them into the source accepted by the attacker and recovering them
  from the distorted sequence.
\end{enumerate}
 
In Section~\ref{ProblemFormulationUnconditional}, we state the precise
formulation of the above problem. In
Section~\ref{MainResultsUnconditional}, we state our main theorem. In
Section~\ref{ConnectionsUnconditional}, we state the connection of the
formulated problem to coding theory, arbitrarily varying channels and
to watermarking with no covertext. In
Section~\ref{ProofsUnconditional}, we prove the theorems stated in
Section~\ref{MainResultsUnconditional} and comment on them in
Section~\ref{Comments}. Section~\ref{MainResultsConditional}
formulates a conditional version of the theorem and it is proved in
Section~\ref{ProofsConditional}. In
Section~\ref{ConnectionsConditional}, we state the relation of this
problem to watermarking. Section~\ref{ContinuousAlphabets} shows how
to generalize to the case of non-finite sources with difference
distortion. Section~\ref{WithMemory}, shows how the results can be
easily extended to stationary ergodic sources that mix appropriately.

Because of space limitations, some of the details in the later
sections are omitted. The full proofs can be found in
\cite{OurDirectConverse}. 

\section{Problem Formulation - Unconditional case}
\label{ProblemFormulationUnconditional}

We start with some notation:

\begin{itemize}

\item 
$\mathcal{X} = \{1, 2, \ldots, |\mathcal{X}|\} \rightarrow$ finite 
set. $\mathcal{X}^\infty$ is the input space.

\item 
$\mathcal{Y} = \{1, 2, \ldots, |\mathcal{Y}|\} \rightarrow$ finite
set. $\mathcal{Y}^\infty$ is the output space.

\item
$p_X \rightarrow$ probability distribution on $\mathcal{X}$.

\item
$X_1^\infty \rightarrow$ iid sequence of random variables, each
$X_i \sim p_X$.

\item
$d: \mathcal{X} \times \mathcal{Y} \rightarrow \mathcal{R}$ is a 
non-negative valued function. We should think of $d(i,j)$ as the 
distortion between $i \in \mathcal{X}$, $j \in \mathcal{Y}$. The focus
is on the average additive distortion on $n$-sequences, 
$ \frac{1}{n} d_n(x_1^n, y_1^n) = \frac{1}{n}\sum_{t=1}^{n} d(x_t, y_t)$. 

\item 
The Attacker is a black box which takes in the input sequence
$x_1^\infty \in \mathcal{X}_1^\infty$ and produces an output
$y_1^\infty \in \mathcal{Y}_1^\infty$. $y_1^\infty$ need not be a
deterministic function of $x_1^\infty$; it can be randomized.

Note that the attacker is, in general, non-causal in the sense that it 
takes in the whole input sequence, looks at it, and produces an
output sequence. The situation that the attacker looks at $x_1^\infty$ 
and produces $y_1^\infty$ is the most general possible. In practice, 
the attacker will only look at finite length sequences and produce an 
output; this is a special case of our definition.

The attacker can also be viewed as a channel. We will use the words attacker 
and attack channel interchangeably.

\item $D$-distortion attacker $\rightarrow$ If the input to the
  attacker is the random variable sequence $X_1^\infty$ (defined above
  - each $X_i$ iid $p_X$), the attacker produces the random variable
  sequence $Y_1^\infty$. This results in a joint probability measure
  on $(X_1^\infty, Y_1^\infty)$. Under this probability measure, there
  should exist some function $f(n)$ with $\lim_{n \to \infty} f(n) = 0$ so that:

\begin{equation} \label{D-distortionAttacker}
\sup_t \Pr \left (\frac{1}{n} \sum_{u = t}^{t+n-1}
 d(X_u, Y_u) > D \right ) \leq f(n) 
\end{equation}

The above equation says that the average distortion caused to long
sequences is bounded by $D$ with high probability, and this
probability $\to$ 1 at least as fast\footnote{No restrictions are made
  on how fast $f(n)$ tends to zero --- just that we know how fast this
  probability goes to zero for this particular family of attackers so
  that we can pick an appropriate block-length for the code.} as
$1-f(n)$ with increasing block lengths $n$ uniformly over at which
time this sliding block\footnote{The purpose of the sliding block is
  merely to reduce notation in stating the condition. All theorems
  will be proved within a single block of length $n$ that is
  sufficiently long on its own. This can be repeated with disjoint
  blocks if a stream of data needs to be transmitted.} is taken
(hence, the name $D$-distortion attacker).

Note that on an individual symbol level, the attacker is essentially
unconstrained --- for any $X_t$, the attacker can distort it
really badly. It is only constrained over very long blocks.

\item $p_X \pm \epsilon$ will denote the set of all probability measures
      $q_X$ on $X$ such that $|q_X(i) - p_X(i)| \leq \epsilon 
      \forall i \in \mathcal{X}$.
      
\end{itemize}

As we can see, the rate-distortion problem when the input sequence is
iid $p_X$ is solved (in the sense of \cite{ControlPartI} by this
attacker for distortion value $D$. The question we want to ask is,
``Can we transmit bits reliably over this attacker in the Shannon
sense, and if yes, at what rates?''

\section{Main results - unconditional case} 
\label{MainResultsUnconditional}

\begin{theorem} \label{MainTheoremUnconditional}
Assuming that there is common randomness available at the transmitter and
the receiver, all rates 
\begin{equation}
R < R_X(D) \triangleq 
 \inf_{\begin{array}{c} X \sim p_X \\ 
               Ed(X,Y) \leq D \end{array}} I(X; Y)
\end{equation}
are achievable over a  $D$-distortion attack channel, and in fact, 
this can be done by using iid $p_X$ random codes.
\end{theorem}
\vspace{0.1in} 
The above theorem says that we can solve the Shannon
communication problem over a $D$-distortion attacker at all rates less
than the rate distortion function, $R_X(D)$. We comment on the need
for common randomness in Section~\ref{Comments} after we prove the
above theorem. 

We also have a converse theorem:
\begin{theorem} \label{Converse}
Rates larger than $R_X(D)$ can in general not be achieved over a 
$D$-distortion attacker.
\end{theorem}
\vspace{0.1in}

After a few comments about this formulation in the next section, it is
proved in the section after next.

\section {Connections to AVCs and Watermarking} 
\label{ConnectionsUnconditional}

We can view the attacker as a non-causal arbitrarily varying channel
(AVC). The AVC is constrained in such a way that it distorts {\em
  most} input sequences to an average distortion less than or equal to
$D$ where ``most'' is according to the iid $p_X$ measure over the
input sequences. The question that we are asking is, ``What is the
capacity of this AVC?'' The foundational papers on AVCs are the papers
by Blackwell, Breiman and Thomasian, \cite{Breiman1,Breiman2}.
\cite{Breiman1} considers the case when the channel is a fixed DMC
coming from a particular set, but unknown. \cite{Breiman2} considers
the case when the channel can vary arbitrarily, but is a DMC at each
time, and comes from a particular set based on past history unlike in
our case where the attack channel at each time does not come from a
particular set, nor is it causal. Stiglitz \cite{StiglitzVarying} has the
same setup as \cite{Breiman2}, but calculates error exponents. Csiszar
and Narayan \cite{csiszar88metric} uses a minimum distance decoding
rule similar to the one that we will use, but it does not consider
AVCs in the form that we do.

To the extent that minimum distance is the relevant idea, this work
can also be considered a generalization of the original formulation of
coding theory in \cite{Hamming} with the distortion measure
generalizing the Hamming distance. In addition, the composition of the
codewords is specified in advance. Fundamentally,
Theorem~\ref{MainTheoremUnconditional} says that every rate-distortion
problem is also associated with a coding theory problem.

This paper's formulation can also be viewed as a watermarking problem
(\cite{Moulin}) with no covertext. The goal is to embed our data in
the input to an attacker that acts within a distortion constraint.
\cite{MerhavPrivate} by Somekh-Baruch and Merhav is the closest to our work.
It allows for non-causal attackers and the definition of attacker is
very similar to ours. But \cite{MerhavPrivate} does not use a minimum
distortion decoding rule --- they use another decoding rule which is
superior in the sense that it achieves the best possible error
exponent. We believe that proofs in \cite{MerhavPrivate}, with slight
modification, should be applicable in our scenario too, but we use a
different decoding rule (a variant of minimum distance decoding) since
it is arguably more natural and achieves capacity. The distinction
between the two papers is more significant in the conditional case.

\section{Proofs - unconditional case}\label{ProofsUnconditional}
We first prove Theorem~\ref{MainTheoremUnconditional} stated in
Section~\ref{MainResultsUnconditional} and show that by using $p_X$
random codes, we can transmit reliably (in the Shannon sense) at all
rates $R < R_X(D)$ over the D-distortion attack channel.
 
\textbf{Codebook Construction}: Generate $2^{nR}$ codewords iid $p_X$. 
This is the codebook, which we denote by $\mathcal{C}$.

\textbf{Decoding}: Fix $\epsilon> 0$. 
Restrict attention to those codewords which are $p_X$-typical, that is,
whose type lies in $p_X \pm \epsilon$ (recall the definition of
$p_X \pm \epsilon$ in Section~\ref{ProblemFormulationUnconditional}: 
all $q_X$ such that $|q_X(i) - p_X(i)| \leq \epsilon \forall i \in 
\mathcal{X}$).

Denote this restricted set of codewords by $\mathcal{C}_R$.

Let $y_1^n$ denote the output of the attacker. If there is a unique 
$p_X$-typical $x_1^n$ in the codebook which is at an average distortion
less than or equal to $D$ from the output sequence, declare that $x_1^n$
was transmitted, else declare error. 
% Mathematically, if
% $\exists ! x_1^n \in \mathcal{C}_R$ such that 
% $ \frac{1}{n}\sum_{t=1}^{n} d(x_t, y_t)
% \leq D$, declare that $x_1^n$ was transmitted, else declare error.

We call our decoding rule the ``$\epsilon$-Nearest Typical Neighbor''
decoding rule. The truly nearest neighbor decoding rule might be a bit
more natural, but it is harder to analyze.

% Our decoding rule is almost the same as the nearest neighbor decoding rule
% except that we are searching for the nearest neighbor among the set of 
% typical codewords.

In what follows,
\begin{itemize}
\item $x_1^n$ denotes the transmitted codeword.
\item $y_1^n$ denotes the received sequence (output of the attacker).
\item $z_1^n$ denotes a $p_X$ typical codeword (that is, $z_1^n
\in \mathcal{C}_R$) such that $z_1^n$ is \textbf{NOT} transmitted. 
\end{itemize}

The error event can be decomposed into 3 parts.
\begin{itemize}

\item 
$E_1 \rightarrow$ transmitted codeword atypical: $x_1^n \notin
\mathcal{C}_R$. 

\item 
$E_2 \rightarrow$ 
Distortion caused by the attacker is not typical:
$\frac{1}{n}\sum_{t=1}^{n} d(x_t, y_t) > D$.

\item 
$E_3 \rightarrow$ a typical codeword which is not transmitted is at an
    average distortion less than or equal to $D$ from the received 
    sequence. Mathematically, $\exists z_1^n \in \mathcal{C}_R$ such that
    $z_1^n$ is not transmitted  and 
    $\frac{1}{n}\sum_{t=1}^{n} d(x_t, y_t) \leq  D$.

\end{itemize}

Clearly, $\Pr(\mbox{error}) \leq \Pr(E_1) + \Pr(E_2) + \Pr(E_3)$.  By
the weak law of large numbers, $\Pr(E_1) \to 0$ as $n \to \infty$.
$\Pr(E_2) \to 0$ as $n \to \infty$ follows by the definition of
$D$-distortion attacker (\ref{D-distortionAttacker}). To upper bound
$\Pr(E_3)$, we do a type-based calculation \cite{csiszarkorner} on the
probability of error for a given received sequence $y_1^n$.

In what follows, it will be helpful to remember that $q$ will always
denote probability measures with \emph{observed} types, whereas $p$
will always denote probability measures with \emph{transmitted} types.
Recall that the received sequence is $y_1^n$. Let the type of $y_1^n$
be $q_Y$, that is, $\forall j \in \mathcal{Y}$, the number of $j$
occurring in $y_1^n$ is $n q_Y(j)$.

Sort the output to place all the $j \in \mathcal{Y}$ together, and
correspondingly shuffle the positions in the codebook's codewords.
This leads to no change in distortion between shuffled codewords and
the sorted received sequence $y_1^n$.

\begin{figure} 
\begin{center}
\psfrag{A1}{$\scriptscriptstyle{nq_{Y}(1)}$}
\psfrag{A2}{$\scriptscriptstyle{nq_Y(j)}$}
\psfrag{A3}{$\scriptscriptstyle{nq_Y(|\mathcal{Y}|)}$}
\psfrag{A4}{$\scriptscriptstyle{nq_Y(j)q_{X|Y}(1|j)}$}
\psfrag{A5}{$\scriptscriptstyle{nq_Y(j)q_{X|Y}(i|j)}$}
\psfrag{A6}{$\scriptscriptstyle{nq_Y(j)q_{X|Y}(|\mathcal{X}||j)}$}
\includegraphics[scale = 0.5]{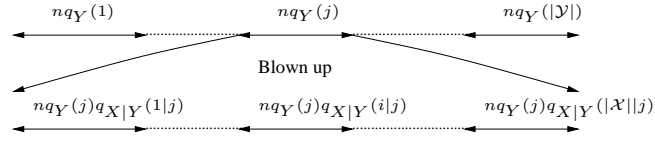}
\end{center}
\caption{The sorted
  received sequence $y_1^n$ and the correspondingly shuffled codeword
  $z_1^n$ illustrating the relevant types.} 
                                 \label{FigureUnconditional}
\end{figure}

Look at a generic shuffled codeword $z_1^n \in \mathcal{C}_R$ which is
not transmitted. Over the chunk of length $nq_Y(j)$, let the type of
the corresponding entries of $z_1^n$ be $q_{X|Y=j}$. (See
Figure \ref{FigureUnconditional})

For the error event $E_3$,

\begin{enumerate}

\item
$z_1^n$ is typical, that is,
\begin{equation}
\sum_{j \in \mathcal{Y}} q_Y(j)q_{X|Y}(i|j) \in p_X \pm \epsilon \forall
i \in \mathcal{X}   
\end{equation}
Denote $\sum_{j \in \mathcal{Y}} q_Y(j)q_{X|Y}(i|j)$ as $q_X(i)$.
Thus, 
\begin{equation}\label{ConstraintOnq_X}
q_X \in p_X \pm \epsilon
\end{equation}

\item
$z_1^n$ is at an average distortion $\leq D$ from the received sequence
$y_1^n$ so 
\begin{equation}
\sum_{i \in \mathcal{X}, j \in \mathcal{Y}}
      q_Y(j)q_{X|Y}(i|j)d(i,j) \leq D
\end{equation}
Denote the distribution $q_Y(j)q{X|Y}(i|j)$ on $\mathcal{X} \times
\mathcal{Y}$ by $q_{X,Y}(i,j)$. Thus,
\begin{equation}\label{ConstraintOnExpectedDistortion}
E_{q_{XY}}d(X,Y) \leq D
\end{equation}
\end{enumerate}
Let us now bound the probability of event $E_3$.

First, the probability that over the chunk of length $nq_Y(j)$, the
corresponding entries of $Z_1^n$ have type $q_{X|Y=j}$ (recall that
$p_X$ is the generating distribution of codeword $Z_1^n$) is given by:
\begin{equation}
\leq 2^{-nq_Y(j)D(q_{X|Y=j}||p_X)}
\end{equation}

Thus, the probability that over the whole block of length $n$, in the chunks 
$nq_Y(j)$, the corresponding entries of $z_1^n$ have type $q_{X|Y=j}$,
for all $j$
\begin{eqnarray}
&\leq& \Pi_{j \in \mathcal{Y}} 2^{-nq_Y(j)D(q_{X|Y=j}||p_X)}\\
&=& 2^{-n \sum_{j \in \mathcal{Y}}q_Y(j)D(q_{X|Y=j}||p_X)} \\
&=& 2^{-n D(q_{XY}||p_X q_Y)}
\end{eqnarray}
It would be helpful to note the positions of where $p$ occur and
where $q$ occur, in the above expression.

To bound the probability that $z_1^n$ is at a distortion $\leq D$ 
from $y_1^n$, we have to sum the above probability over all possible
types $q_{X|Y=j}, 1 \leq j \leq |\mathcal{Y}|$ such that conditions 
1 and 2 above (equivalently,  (\ref{ConstraintOnq_X}) and 
(\ref{ConstraintOnExpectedDistortion})) are satisfied.

Number of $q_{X|Y=j}$ types $\leq (n+1)^{|\mathcal{X}||\mathcal{Y}|}$.
Also recall that number of non-transmitted codewords $|\mathcal{C}_R|
\leq 2^{nR}$.

Putting all this together and using the union bound, 
\begin{eqnarray}
&&\Pr(E_3|\ \mbox{type of} \ y_1^n \ \mbox{is}\ q_Y) \\ \nonumber
&& \hspace{0.5cm}\leq    (n+1)^{|\mathcal{X}||\mathcal{Y}|} 2^{nR} 
          2^{-n \inf _ {q_{XY}\in \mathcal{S}} D(q_{XY}||p_X q_Y) }
\end{eqnarray}
where $\mathcal{S}$ denotes the set of types satisfying conditions 1 and 2
(equivalently, (\ref{ConstraintOnq_X}) and
(\ref{ConstraintOnExpectedDistortion})), and is
\begin{equation}
\mathcal{S} =  \left \{ q_{XY}: 
       \begin{array}{l} q_X \in p_X \pm \epsilon \\
                      E_{q_{XY}}d(X,Y) \leq D \\
                      q_Y \ \mbox{fixed}
       \end{array} \right \}
\end{equation}

Now, $q_Y$, the type of the received sequence $y_1^n$ is arbitrary.
Thus, an easy way to bound $\Pr(E_3)$ is to just remove the 
$q_Y$ fixed condition from the above definition of $\mathcal{S}$.

Thus finally,
\begin{equation}
\Pr(E_3) \leq  
       (n+1)^{|\mathcal{X}|(|\mathcal{Y}|+1)} 2^{nR} 
          2^{-n \inf_{q_{XY}\in \mathcal{T}} D(q_{XY}||p_X q_Y) }
\end{equation}
where $\mathcal{T}$ is the set
\begin{equation}
\mathcal{T} =  \left \{ q_{XY}: 
       \begin{array}{l} q_X \in p_X \pm \epsilon \\
                      E_{q_{XY}}d(X,Y) \leq D \\
       \end{array} \right \}
\end{equation}
The only difference between the sets $\mathcal{S}$ and $\mathcal{T}$ is that
the $q_y$ fixed condition which exists in $\mathcal{S}$ has been removed
in $\mathcal{T}$.

Since $(n+1)^{|\mathcal{X}|(|\mathcal{Y}|+1)}$ is a polynomial, 
$\Pr(E_3) \to  0$ as $n \to \infty$ if 
\begin{equation}
R < \inf_ {\begin{array}{c} q_X \in p_X \pm \epsilon \\
         Ed_{q_{XY}}(X,Y) \leq D \end{array}} D(q_{XY}||p_X q_Y) \\
\end{equation}

Thus to prove Theorem~\ref{MainTheoremUnconditional}, it suffices to
prove that 
\begin{eqnarray}
 \Theta_1 &\triangleq&
\lim_{\epsilon \to 0}
 \inf_ {\begin{array}{c} q_X \in p_X \pm \epsilon \\
         E_{q_{XY}}d(X,Y) \leq D \end{array}} \hspace{-1cm} D(q_{XY}||p_X q_Y) \\ \nonumber
&=& R_X(D)
= \inf_{\begin{array}{c} X \sim p_X  \\
                      Ed(X,Y) \leq D \end{array}} \hspace{-.3cm} 
                                                       I(X;Y) \\ \nonumber
&=& \hspace{-0.5cm}\inf_{\begin{array}{c} p_X \ \mbox{fixed} \\
                                  p_Y \ \mbox{can vary}\\          
               E_{p_{XY}}d(X,Y) \leq D \end{array}} \hspace{-1cm}D(p_{XY}||p_X p_Y)
\triangleq \Theta_2
\end{eqnarray}    
The main difference between $\Theta_1$ and $\Theta_2$ (note the definitions
of $\Theta_1$ and $\Theta_2$ in the above equation) is that:
\begin{itemize}

\item
In $\Theta_1$, we have $D(q_{XY}||p_X q_Y)$; $q_X \in p_X \pm \epsilon$

\item
In $\Theta_2$, we have $D(p_{XY}||p_X p_Y)$
\end{itemize}
It is clear that $\Theta_1$ has ``more freedom'' and hence, 
$\Theta_1 \leq \Theta_2$.

All we need to prove is that $\Theta_1 \geq \Theta_2$.

This we do with a simple trick:
\begin{eqnarray}
D(q_{XY}||p_X q_Y) = D(q_{X}||p_X) + D(q_{XY}||q_X q_Y)  \\ \nonumber
   \hspace{3cm}\geq D(q_{XY}||q_X q_Y)
\end{eqnarray}
Thus,
\begin{equation}
\Theta_1 \geq \lim_{\epsilon \to 0} 
        \inf_ {\begin{array}{c} q_X \in p_X \pm \epsilon \\
         E_{q_{XY}}d(X,Y) \leq D \end{array}} D(q_{XY}||q_X q_Y) \\
\end{equation}
So we only need to prove that
\begin{eqnarray}
 \lim_{\epsilon \to 0} 
        \inf_ {\begin{array}{c} q_X \in p_X \pm \epsilon \\
         E_{q_{XY}}d(X,Y) \leq D \end{array}} D(q_{XY}||q_X q_Y) \\ \nonumber
\geq
\lim_{\epsilon \to 0} 
        \inf_ {\begin{array}{c} p_X \ \mbox{fixed}\\
         Ed(X,Y) \leq D \end{array}} D(p_{XY}||p_X p_Y)
\end{eqnarray}
This holds with equality, and follows from the continuity of the rate
distortion function $R_X(D)$ in $p_X$ and proves the direct theorem.

The sequence of choosing $n,\epsilon$ depending on the rate $R <
R_X(D)$ and probability of error $p_e$ is:
\begin{enumerate}
\item
Choose $\epsilon$ small enough so that $R < \inf_{X \in p_X \pm \epsilon}
R_X(D)$. 
% The reason for this is that we cannot transmit at rates larger
% than $\inf_{X \in p_X \pm \epsilon}R_X(D)$ with 
% ``$\epsilon$-Nearest typical decoding''

\item
Choose $n$ large enough so that the total probability of error
from the events $E_1$, $E_2$ and $E_3$ adds up to a value less than
$p_e$.
\end{enumerate}

We now sketch the proof of the converse theorem,
Theorem~\ref{Converse}, that is, in general, we cannot transmit at
rates larger than $R_X(D)$ over a $D$-distortion attacker. Another way
of stating this is that if one tries to transmit at rates larger than
$R_X(D)$, there is a $D$-distortion attacker such that we cannot
transmit reliably over this attacker.

First, consider the case that we are restricted to using
iid $p_X$ random codes;
we will remove this restriction later.

Let the rate at which we want to transmit,
$R = R_X(D - \alpha) > R_X(D)$ for some $\alpha > 0$.

We will show that there is a D-distortion attacker which is a DMC for which
error probability $\nrightarrow$ 0.

Look at all DMCs that produce an average distortion of 
$(D - \frac{\alpha}{2})$ between the input and output when input is 
$p_X$ distributed. 
% The capacity of the worst channel among these channels
% when restricted to $p_X$ random codes is the minimum possible mutual 
% information between input and output when restricting to DMCs which produce
% an average distortion of $(D - \frac{\alpha}{2})$ between the input and
% the output when the input distribution is restricted to $p_X$. That is,
\begin{equation}
C_{\mbox{worst}} =  
\inf_ {\begin{array}{c} X \sim p_X \\
         Ed(X,Y) \leq (D - \frac{\alpha}{2}) \end{array}} I(X;Y)
\end{equation}
But this value is precisely $R_X(D - \frac{\alpha}{2})$. Also,
any DMC that produces an average distortion of $(D - \frac{\alpha}{2})$ is 
a $D$-distortion attacker (follows from the weak law of large numbers). Thus,
we have exhibited a DMC which is a $D$-distortion attacker and over which, 
we cannot reliably at rates larger than $R_X(D - \frac{\alpha}{2}) < 
R_X(D - \alpha) = R$.

To remove the assumption that we have to use $p_X$ random codes,
consider the following attacker: 

Fix $\epsilon > 0$. The attacker looks at inputs of length $n$ and if
the input is not $p_X$ typical (that is, the empirical type does not
lie in $p_X \pm \epsilon$), the attacker will produce junk output, say
the all $1$ sequence, whereas if the input sequence is $p_X$-typical,
the attacker will act like the above DMC. The attacker needs to keep
increasing the length of sequences which it looks at and attacks, and
correspondingly decrease $\epsilon$. It is intuitively clear that if a
codebook is chosen with a codeword which is not $p_X$-typical, the
output of the attacker will give no positive rate information about
what was transmitted, and hence, the encoder can not use such
codewords to transmit reliably at rates larger than $R_X(D)$.

% This argument can be made rigorous, but we omit it in the interests of
% space.

\section{Comments on the proof} \label{Comments}

% \item One need not be worried about the attacker looking at infinite length
% sequences and producing output. For codebook formation, we just need to choose
% an $n$ large enough so that the attacker causes 
% a distortion larger than $D$ with probability less 
% than what we can tolerate, and we are fine.
% 
% \item It is possible that just minimum distance decoding will work 
% rather than doing the ''$\epsilon$-Nearest typical decoding''; we don't know.
% 
% The event $E_3$ requires further comment:

% We generate $2^{nR}$ codewords iid $p_X$. We draw balls of radius $D$
% around them. Having $\Pr(E_3)$ small is equivalent to the question that
% what is the maximum $R$ that we can have for which the union of the above
% balls will not cover the received sequence (with high probability), where 
% the received sequence is arbitrary but is generated independently of
% the ($2^{nR}$ - 1) non-transmitted codewords. 
% 
% In Rate-Distortion theory, we generate $2^{nR}$ codewords iid $p_{\hat
%   X}$, where $p_{\hat X}$ is arbitrary, but we can choose it. We draw
% balls of radius $D$ around them. Having a small probability of
% distortion larger than $D$ is equivalent to the question that what is
% the minimum $R$ that we need to have for which the union of the above
% balls will cover (with high probability) the iid $p_X$ sequence 
% that we need to compress.
% 
% The two look exactly the opposite and there is
% a striking duality between the two; it is not accidental that
% we get the same answer $R = R_X(D)$ for both questions above.

If one compares the proofs of Shannon's channel coding theorem and the
above, the two are quite similar in the error calculation for the
event $E_3$, but there is one difference. In Shannon's theorem,
proving that the average error probability over the ensemble of codes
$\to$ 0 implies that there exists a codebook for which the error
probability $\to$ 0 for every single message. This is not immediately
true in our case because the attacker can use different strategies
over different blocks.
 
Furthermore, if we were to use the same codebook over and over again,
the input would no longer look iid $p_x$ on very long sequences and
the attacker would be free to just drive us to zero.  Thus, the
codebook has to be generated at least somewhat independently in each
block of length $n$. This is where we use the assumption that there is
common randomness available --- using this common randomness, the
transmitter and the receiver can generate the codebook again and
again, independently.

However, the code as given requires an exponentially large amount of
common randomness. This can easily be reduced to a polynomial (in the
block-length $n$) amount of common randomness by using the following
tricks:(details in \cite{OurDirectConverse})
  \begin{itemize}
   \item Simulate in advance whether the input block will be
         $\epsilon$-typical or not. (Can use $O(\log n)$ bits) If it
         is atypical, just declare error no matter what message was
         sent. 

       \item Make slight modifications to the proof to instead show
         the existence of deterministic codebooks with input types
         like $p_x \pm \epsilon$ that can be list-decoded to some
         possibly large, but constant, list-size $l$ when facing a
         worst-case attacker inducing a distance $D$. This is done by
         patching the above proof with arguments analogous to those
         for Theorem 5.1 in \cite{GVThesis}. The additional trick is
         just noticing that $I(X;Y) = H(Y) - H(Y|X)$ and that
         $2^{nH(Y)}$ is essentially the total number of output
         sequences\footnote{Rather than computing the probability of
           error, we are computing the expected number of $D$-balls
           that have at least $l+1$ codewords in them. For a given
           $l+1$ codeword positions, this is just the existing
           probability of collision raised to the $l+1$ power times
           the number of possible $D$-balls. The total number of such
           combinations is also no more than $2^{nR(l+1)}$.} of type $q_Y$.
         When $l$ is large enough, $\frac{l}{l+1}H(Y) - H(Y|X)$ is as
         close as desired\footnote{And so the expected total number of
           collisions is as small as we want and so there exists at
           least one deterministic codebook that has no such
           collisions at the $l$-list level.} to $R_X(D)$.

   \item Once the deterministic codes are constant composition, a
         random permutation of the indices will make each of them
         behave as though they were drawn from the original iid $p_x$
         distribution conditioned on the empirical type being typical.
         This takes $O(n\log(n))$ commonly-random bits.

   \item By using the code at a rate slightly less than the rate of
         the code, the message can be padded with a randomly chosen
         hash of the true message. This takes at most another $O(n)$
         commonly-random bits and allows the decoder to uniquely
         disambiguate the decoded lists with high probability by just
         rejecting messages whose hashes do not match up correctly.  
  \end{itemize}    

\section{Theorem - conditional Case}\label{MainResultsConditional}

Until now, we assumed that the input to the attacker should be a
$p_X$-iid sequence. Now, consider the case that the input is still an
independently generated sequence but the distribution of $X_i$ depends
on an iid random variable sequence $V_1^\infty$ that is revealed
non-causally to all parties.

We state some notation to add to the notation previously.

\begin{itemize}

\item
$\mathcal{V} \rightarrow = \{1, 2, \ldots |\mathcal{V}|\}$ is a finite set.
A generic element of $\mathcal{V}$ will be denoted by $s$.

\item
$p_V \rightarrow$ probability distribution on $\mathcal{V}$.

\item $V_1^\infty \rightarrow$ iid sequence of random variables
  generated $p_V$. In watermarking terms, this can be thought of as
  the ``cover-story.'' We will talk about relations to watermarking in
  Section~\ref{ConnectionsConditional}.

\item $p_{X|V=s} \rightarrow$ If $V_i = s$, $X_i$ is generated according
to the distribution  $p_{X|V=s}$, but independently of other $X_j$. The
joint distribution on $(V_i, X_i)$ will be denoted by $p_{VX}$

\item Attacker $\rightarrow$
\textbf{We assume that} $V_i$ \textbf {is known noncausally to the
  encoder, decoder and the attacker}.
% The attacker looks at the sequence $(v_1^\infty, x_1^\infty)$ and generates
% the output sequence $y_1^\infty$.

% \item $D$-distortion attacker $\rightarrow$ An attacker produces a joint
% probability measure on $(V_1^\infty, X_1^\infty, Y_1^\infty)$. If 
% under this probability measure, (\ref{D-distortionAttacker})
% holds, that is,
% \begin{equation} 
% \sup_t \Pr \left (\frac{1}{n} \sum_{u = t}^{t+n-1}
%  d(X_u, Y_u) > D \right ) \to 0\  \mbox{as $n$ } \to \infty 
% \end{equation}
% then we say that the attacker is a $D$-distortion attacker. This is the 
% same definition that we gave previously and has the same intuitive meaning.
\end{itemize}

The next theorem is a conditional version of the inverse rate-distortion
theorem, Theorem \ref{MainTheoremUnconditional}.

\begin{theorem} \label{MainTheoremConditional}
Assuming that there is common randomness available at the transmitter and
the receiver, all rates
\begin{equation}
R < R_{X|V}(D) \triangleq 
   \inf_{\begin{array}{c} (V,X) \sim p_{VX} \\  Ed(X,Y) \leq D
               \end{array}} I(X; Y|V)
\end{equation}
are achievable over a  $D$-distortion attack channel, and in fact, 
this can be done by using iid $p_{X|V}$ random codes.
\end{theorem}
\vspace{0.1in}
We omit a converse theorem though the same arguments as above would
give one.

\section{Proofs - conditional case}\label{ProofsConditional}
The proof is very similar to the proof of the theorem in the
unconditional case. Recall that $V_1^\infty$ is known to the
transmitter, receiver, and attacker.

\textbf{Codebook Construction}: Generate $2^{nR}$ codewords iid $p_{X|V}$. 
This is the codebook, which we denote by $\mathcal{C}$.

\textbf{Decoding}: Fix $\epsilon > 0$. 
Restrict attention to those codewords $x_1^n$ such that $(v_1^n, x_1^n)$ 
is $p_{VX}$ typical, that is,
whose type lies in $p_{VX} \pm \epsilon$.

Denote this restricted set of codewords by $\mathcal{C}_R$.

Note that if $v_1^n$ is not typical, $C_R$ will be empty. Thus:
\begin{itemize}

\item
The definition of $\mathcal{C}_R$ implicitly assumes an error if $v_1^n$
is not strongly typical.

\item $\mathcal{C}_R$ depends on $v_1^n$, that is, the codewords of 
$\mathcal{C}$ which lie in $\mathcal{C}_R$ are different for different
$v_1^n$.
\end{itemize}

Let $y_1^n$ denote the output of the attacker. If there is a unique
$x_1^n$ in the restricted codebook which is at an average distortion
less than or equal to $D$ from the output sequence, declare that
$x_1^n$ was transmitted, else declare error. We call this the
``$\epsilon$-Nearest Conditionally Typical Neighbor'' decoding rule.

In what follows, $z_1^n$ will denote a non-transmitted codeword as
before. As in the unconditional case, the error event
consists of 3 parts: 

%($x_1^n
%y_1^n, z_1^n$ will be the same as in the unconditional case).

%\begin{itemize}

%\item
%$v_1^n \rightarrow$ $V_1^n$ sequence realized.

%\item
%$x_1^n \rightarrow$ transmitted codeword.

%\item
%$y_1^n \rightarrow$ received codeword.

%\item
%$z_1^n \rightarrow$ $z_1^n\in \mathcal{C}_R$, but $z_1^n$ is not transmitted.

%\end{itemize}
%$x_1^n,y_1^n, z_1^n$ are defined in the same way as in the unconditional case.

\begin{itemize}

\item 
$E_1 \rightarrow$ $(v_1^n, x_1^n)$ is not typical. This is a slight modification
of $E_1$ in the unconditional case.

\item 
$E_2 \rightarrow$ 
Distortion caused by the attacker is not typical, that is,
transmitted codeword is at an average distortion
larger than $D$ from the received sequence. Mathematically,
$\frac{1}{n}\sum_{t=1}^{n} d(x_t, y_t) > D$. This is exactly the 
same as in the unconditional case.

\item $E_3 \rightarrow$ a typical codeword which was not transmitted
  is at an average distortion less than or equal to $D$ from the
  received sequence. This is exactly the same as in the
  unconditional case.

\end{itemize}

$\Pr$(error) $\leq$ $\Pr(E_1)+ \Pr(E_2)+ \Pr(E_3)$.
$\Pr(E_1), \Pr(E_2) \to 0$ as in the unconditional case. 

All we need to do is to upper bound $\Pr(E_3)$. As before, we do a
method-of-types calculation on the probability of possible $z_1^n$
that will cause an error for a given received sequence $y_1^n$.

The only essential difference between this proof and in the proof of
the unconditional case is that we first do a sorting based on $V$ and
then proceed exactly the same as before, that is, do a sorting based
on $Y$ and then do a sorting based on $X$.

\begin{figure} 
\begin{center}
\psfrag{A1}{$\scriptscriptstyle{nq_{V}(1)}$}
\psfrag{A2}{$\scriptscriptstyle{nq_V(s)}$}
\psfrag{A3}{$\scriptscriptstyle{nq_V(|\mathcal{V}|)}$}

\psfrag{A4}{$\scriptscriptstyle{nq_V(s)q_{Y|V}(1|s)}$}
\psfrag{A5}{$\scriptscriptstyle{nq_V(s)q_{Y|V}(j|s)}$}
\psfrag{A6}{$\scriptscriptstyle{nq_V(s)q_{Y|V}(\mathcal{|\mathcal{Y}|}|s)}$}

\psfrag{A7}{$\scriptscriptstyle{nq_V(s)q_{Y|V}(j|s)
                          q_{X|VY}(i|s,j)}$}

\includegraphics[scale = 0.5]{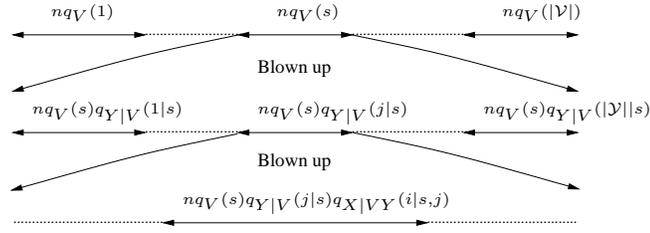} 
\end{center}
\caption{The various types
  illustrated in the conditional rate-distortion case.} \label{FigureConditional}

\end{figure}

Let the type of $v_1^n$ look like $q_V$. Sort, so that all $t$ such
that $V_t = s$ are together. Over the subsequence where $V_t = s$, let
the type of the output produced by the attacker be $q_{Y|V=s}$. Again,
do a sub-sorting such that all $Y_t = j$ are together in each
subsequence of $V_t = s$. In this $(V_t = s, Y_t = j)$ subsequence,
let the type of the subsequence of $z_1^n$ (recall - $z_1^n$ is a
codeword which is NOT transmitted) look like $q_{X|Y=j, V=s}$. See
Figure \ref{FigureConditional}. 
 
We now do the $\Pr(E_3)$ calculation.

First restrict attention to the subsequence $V_t = s$. Over this subsequence,
we do exactly what we did in the unconditional case. It follows from
the proof of the unconditional case that the probability that $Z_1^n$
looks like $q_{X|V=s, Y=j}$ given that the $y_1^n$ subsequence type
looks like $q_{Y|V=s}$ is
\begin{equation}
\leq 2^{-nq_V(s)D(q_{XY|V=s}||p_{X|V=s}q_{Y|V=s})}
\end{equation}

The probability that over the whole sequence, the $Z_1^n$ type is
$q_{X|V,Y}$ given that the $Y$ type is $q_{Y|V}$
\begin{eqnarray}
&\leq& 2^{-n\sum_{s \in \mathcal{V}}q_V(s)D(q_{XY|V=s}||p_{X|V=s}q_{Y|V=s})}\\
&=& D(q_{XY|V}||p_{X|V}q_{Y|V}|q_V)
\end{eqnarray}
There are a polynomial number of $q_{VXY}$ types,
$\leq (n+1)^{|\mathcal{V}| |\mathcal{X}| |\mathcal{Y}| }$
and by argument similar to that in the unconditional case,
\begin{eqnarray}
&&Pr(E_3) \leq \\ \nonumber 
&&  2^{nR} (n+1)^{|\mathcal{V}| | \mathcal{X}| (|\mathcal{Y}|+1)}
     2^{-n \inf_{q_{VXY} \in \mathcal{R}}
           D(q_{XY|V}||p_{X|V}q_{Y|V}|q_V)}
\end{eqnarray}
where the set $\mathcal{R}$ over which the above infimum is taken is:

\begin{enumerate}

\item
$(v_1^n, z_1^n)$ is typical, that is,
$q_{VX} \in p_{VX} \pm \epsilon$.

\item
$z_1^n$ is at an average distortion $\leq D$ from the received sequence
$y_1^n$, that is, 
$E_{q_{VXY}}d(X,Y) \leq D$

\end{enumerate}

Thus, 
\begin{equation}
\mathcal{R} =  \left \{ q_{VXY}: 
       \begin{array}{l} q_{VX} \in p_{VX} \pm \epsilon \\
                      E_{q_{VXY}}d(X,Y) \leq D 
       \end{array} \right \}
\end{equation}

It follows that we only need to prove that 
\begin{eqnarray} \label{ConditionalEquation}
\lim_{\epsilon \to 0}
 \inf_ {\begin{array}{c} q_{VX} \in p_{VX} \pm \epsilon \\
         E_{q_{VXY}}d(X,Y) \leq D \end{array}} 
            D(q_{XY|V}||p_{X|V}q_{Y|V}|q_V) \\ \nonumber
= R_{X|V}(D)
= \inf_{\begin{array}{c} (V,X) \sim p_{VX}  \\
                      Ed(X,Y) \leq D \end{array}} I(X;Y|V) \\ \nonumber
= \inf_{\begin{array}{c} p_{VX} \ \mbox{fixed} \\          
               Ed_{p_{VXY}}(X,Y) \leq D \end{array}} 
                   D(p_{XY|V}||p_{X|V}p_{Y|V}|p_V)
\end{eqnarray}    
The proof of this follows in almost the same way as in the
unconditional case, just that we have to use the continuity of
$R_{X|V}(D)$ in $p_{VX}$ (in the unconditional case, we had used the
continuity of $R_X(D)$ in $p_X$).

This proves the conditional theorem, Theorem~\ref{MainTheoremConditional}.

% The sequence of choosing $n$ and $\epsilon$ depending on the rate $R <
% R_{X|V}(D)$ and probability of error $p_e$ is:
% \begin{enumerate}
% \item
% Choose $\epsilon$ small enough so that $R < \inf_{(V,X) \in 
% p_{VX} \pm \epsilon}
% R_{V|X}(D)$. 
% 
% \item
% Choose $n$ large enough so that the total probability of error
% from the events $E_1$, $E_2$ and $E_3$ adds up to a value less than
% $p_e$.
% 
% \end{enumerate}

\section{Relation to Watermarking} \label{ConnectionsConditional} We
can view this conditional problem as a watermarking problem with a
coverstory\footnote{To distinguish it from the ``covertext'' in
  traditional watermarking}. In watermarking, the user is allowed to
make some tolerable level of distortion to the covertext. We have a
restriction of another kind, that is, if the coverstory entry is $s$,
the input distribution should be $p_{X|V=s}$. Also, in watermarking,
the covertext is not known to the attacker.\footnote{Since otherwise,
  presumably the attacker could just replace the input with the
  covertext itself. The same is not true if it is considered as a
  coverstory.} We assume that the covertext is known to the attacker.
If one looks at (38) in the paper of Somekh-Baruch and Merhav
\cite{MerhavPrivate}, this is the reason for the Markov Chain condition $U
\rightarrow X \rightarrow Y$.  We do not have the Markov Chain
condition $V \rightarrow X \rightarrow Y$ because the covertext $V$ is
known to the attacker.

\section{Continuous alphabets}\label{ContinuousAlphabets}
In this section, we consider the case when $\mathcal{X}$, $\mathcal{Y}$ 
and $\mathcal{V}$ are not necessarily finite discrete alphabets. We
divide the problem into 6 cases:
\begin{enumerate}

\item
$\mathcal{X}$ finite,  $\mathcal{Y}$ finite,  $\mathcal{V}$ not there.

\item
$\mathcal{X}$ finite   $\mathcal{Y}$ finite,  $ \mathcal{V}$ finite.

\item
$\mathcal{X}$ non-finite,  $\mathcal{Y}$ non-finite,  $\mathcal{V}$ not there.

\item
$\mathcal{X}$ non-finite, $\mathcal{Y}$ non-finite,  $\mathcal{V}$ finite.

\item
$\mathcal{X}$ finite,  $\mathcal{Y}$ finite,  $\mathcal{V}$ non-finite.

\item
$\mathcal{X}$ non-finite,  $\mathcal{Y}$ non-finite,  $\mathcal{V}$ non-finite.
\end{enumerate}

We will refer to these as Cases 1 through 6. Case 1 is the
unconditional case covered in Theorem~\ref{MainTheoremUnconditional},
Case 2 is the conditional case covered in
Theorem~\ref{MainTheoremConditional}. We now go on to the rest. The
proofs will be based on quantization of the above sets and using ideas
from the proofs of Theorem~\ref{MainTheoremUnconditional} and
\ref{MainTheoremConditional}.

For Case 3, we need to prove that rates $< R_X(D)$ are achievable and for
Cases 4,5,6, we need to prove that rates $< R_{X|V}(D)$ are achievable.

\begin{figure}  
\begin{center}
\includegraphics[scale = 0.5]{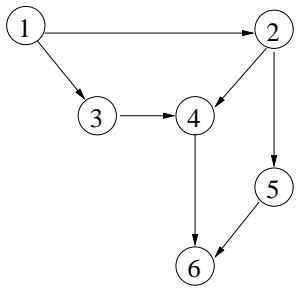}
\end{center}
\caption{Dependence graph for the proofs of the various cases}
\label{Dependence}
\end{figure}

Figure~\ref{Dependence} is a dependency graph of which proofs depend on which.

%%%%%%%%%%%%%%%%%%%%%%%%%%%%%put figure%%%%%%%%%%%%%%%%%%%%%%%%%%%%%%%%%%%

\subsection{Compact support}
We first tackle Case 3, that is, $\mathcal{X}, \mathcal{Y}$ are
non-finite sets, and there is no $\mathcal{V}$. We first assume that
$\mathcal{X}$ and $\mathcal{Y}$ are bounded subsets of
$\mathcal{R}^\gamma$, for some positive integer $\gamma$. The case of
unbounded support is addressed later.

We first state some notation:

\begin{itemize}

\item 
$\mathcal{X}, \mathcal{Y} \rightarrow$ \textbf{bounded} subsets of 
$\mathcal{R}^\gamma$.

\item 
$a \rightarrow$ generic point in $\mathcal{X}$. We do not use $x$ 
because of potential confusion with the transmitted sequence.

\item $b \rightarrow$ generic point in $\mathcal{Y}$. We do not use
  $y$ because of potential confusion with the received sequence.

\item $d: \mathcal{R^\gamma} \times \mathcal{R}^\gamma \rightarrow
  \mathcal{R}$ is a difference distortion measure which is assumed to
  be \textbf{uniformly continuous} with respect to the Euclidean
  metric.

\item $D$-distortion attacker $\rightarrow$ Same as before. If the
  input to the attacker is the random variable sequence $X_1^\infty$ (
  $X_i$ iid $p_X$) , the attacker produces the random variable
  sequence $Y_1^\infty$. This results in a joint probability measure
  on $(X_1^\infty, Y_1^\infty)$. Under this probability measure,

\begin{equation}
\sup_t \Pr \left (\frac{1}{n} \sum_{u = t}^{t+n-1}
 d(X_u, Y_u) > D \right ) \to 0\  \mbox{as $n$ } \to \infty 
\end{equation}

\item $\mathcal{X}_\Delta, \mathcal{Y}_\Delta \rightarrow$
  $\Delta$-hypercube grid quantization of $\mathcal{X}, \mathcal{Y}$
  respectively. The boundary of the hypercube can be put in any of the
  adjoining sets but not both. The quantization point is taken as the
  center of the hypercube.

\item $a_\Delta \rightarrow$ Generic point of $\mathcal{X}_\Delta$.
  $a_\Delta \in \mathcal{X}_\Delta$ is obtained by quantizing $a \in
  \mathcal{X}$.

\item $b_\Delta \rightarrow$ Generic point of $\mathcal{Y}_\Delta$.
  $b_\Delta \in \mathcal{Y}_\Delta$ is obtained by quantizing $b \in
  \mathcal{Y}$.

\item $p_{X_\Delta} \rightarrow$ Probability distribution on
  $\mathcal{X}_\Delta$ obtained from the distribution $p_X$ on
  $\mathcal{X}$ in the obvious way.

\end{itemize}

Note that since the difference distortion function is uniformly
continuous and $\mathcal{X}, \mathcal{Y}$ are bounded, $\frac{1}{n}
\sum_{t = 1}^{n}d(x_{t\Delta}, y_{t\Delta}) \leq \frac{1}{n} \sum_{t =
  1}^{n}d(x_{t}, y_{t}) + g(\Delta) \forall (x_1^n, y_1^n) \in
\mathcal{X}^n \times \mathcal{Y}^n$ where $g(\Delta) \to 0$ as $\Delta
\to 0$.

It follows that under the distribution governing $(X_{1\Delta}^n,
Y_{1\Delta}^n)$ under the $D$-distortion attacker, 
\begin{equation}
\sup_t \Pr \left(\frac{1}{n} \sum_{u = t}^{t+n-1}
 d(X_{u\Delta}, Y_{u\Delta}) > D + g(\Delta) \right) 
                          \to 0\  \mbox{as }n \to \infty 
\end{equation}
If we work in the quantized world, this suggests what the decoding
rule should be.

\textbf{Codebook Construction}: Generate $2^{nR}$ codewords iid $p_X$. 
This is the codebook $\mathcal{C}$. Let $\mathcal{C}_\Delta$ denote
the quantized codebook obtained by quantizing each codeword. 

\textbf{Decoding}: Fix $\epsilon > 0$. 

Restrict attention to those quantized codewords which are 
$p_{X_\Delta}$-typical. Denote this restricted set of 
quantized codewords by $\mathcal{C}_R^\Delta$.

Let $y_{1\Delta}^n $ denote the quantized output of attacker. 
If there is a unique 
$p_{X_{\Delta}}$-typical quantized codeword $x_{1\Delta}^n$
which is at an average distortion
less than or equal to $D + g(\Delta)$ (\textbf{note the change} 
$D + g(\Delta)$ \textbf{instead of} $D$)
from the output sequence, declare that $x_1^n$
was transmitted, else declare error. Mathematically, if
$\exists ! x_1^n \in \mathcal{C}_R$ such that 
$ \frac{1}{n}\sum_{t=1}^{n} d(x_t, y_t)
\leq D + g(\Delta)$, declare that $x_1^n$ was transmitted, else declare error.

This decoding rule has reduced the problem to Case 1(finite
$\mathcal{X}$ and $\mathcal{Y}$), and we can use results from there.
Thus, we can transmit at rates $R < R_{X_\Delta}(D + g(\Delta))$ using
this decoding rule.  It can be shown using the appropriate continuity
arguments that $\lim_{\Delta \to 0} R_{X_\Delta}(D + g(\Delta)) =
R_X(D)$. This proves that we can transmit at all rates $< R_X(D)$.

The sequence in which $n, \Delta, \epsilon$ need to be chosen 
depending on the desired rate $R < R_X(D)$ and the error probability
$p_e$ is: 
\begin{enumerate}

\item 
Choose $\Delta$ small enough so that $R < R_{X_\Delta}(D + g(\Delta))$.

\item Choose $\epsilon$ small enough so that 
$R < \inf_{X_\Delta \in p_{X_\Delta} \pm \epsilon}R_{X_\Delta}(D + g(\Delta))$. 

\item Choose $n$ large enough so that the sum of error probabilities
  of events $E_1, E_2, E_3$ is less than $p_e$. 
\end{enumerate}

Case 4, where $\mathcal{X}, \mathcal{Y}$ are non-finite while the
``coverstory'' $\mathcal{V}$ is finite, is proved in exactly the same
way --- by quantizing $X,Y$ finely enough. 

% Again, quantize $\mathcal{X}$ and $\mathcal{Y}$ as in Case 3. 
% Construct a decoding rule which will ``mix'' the decoding rules of Case
% 2 ($\mathcal{X}$ finite   $\mathcal{Y}$ finite,  $ \mathcal{V}$ finite).
% and Case 3 (above, that is, $\mathcal{X}$ non-finite, $\mathcal{Y}$ 
% non-finite,  $ \mathcal{V}$ not there) in the obvious way. We omit a formal
% proof. It follows that we can transmit at rates $R < R_{X|V}(D)$.

Next we consider Case 5, that is,
$\mathcal{X}, \mathcal{Y}$ are finite and  $\mathcal{V}$ is non-finite.
We assume that $\mathcal{V}$ is a bounded subset of $\mathcal{R}^\eta$
for some positive integer $\eta$.

We introduce some notation regarding $\mathcal{V}$. 

\begin{itemize}

\item
$\mathcal{V} \rightarrow$ bounded subset of $\mathcal{R}^\eta$.

\item 
$s \rightarrow$ generic element of $\mathcal{V}$.

\item
$\mathcal{V}_{\Delta'} \rightarrow$ $\Delta'$ hypercube quantization
of $\mathcal{V}$. The boundary of the hypercube can be put in any of the
adjoining sets. Quantization is taken as the center of the hypercube.
We use $\Delta'$ instead of $\Delta$ because we use $\Delta$ for quantizing
$\mathcal{X}$ and $\mathcal{Y}$.

\item 
$s_{\Delta'} \rightarrow$ Generic point of $\mathcal{V}_{\Delta'}$.
$s_{\Delta'}$ in $\mathcal{V}_{\Delta'}$ is got by quantizing 
$s$ in $\mathcal{V}$.

\item
$\mathcal{S}_{\Delta'} \rightarrow$ quantization region
(hypercube) of $\mathcal{V}$ containing
the point $s_{\Delta'} \in \mathcal{V}_{\Delta'}$.

\item $p_{V_{\Delta'}} \rightarrow$ probability distribution on
$\mathcal{V}_{\Delta'}$ got from $p_{V}$ on $\mathcal{V}$ in the obvious 
way.
\end{itemize}

What is not obvious, though, is how to define $p_{X|V_{\Delta'}}$. We 
need to make definitions in such a way that we can do probability of error
calculations for the event $E_3$ (the other two events, $E_1$ and $E_2$ will
be trivial as usual).
% 
% We make the following definitions:
% 
\begin{eqnarray}
p_{X|V_{\Delta'}=s_{\Delta'}}^{\sup} &=& \sup_{s \in  \mathcal{S}_{\Delta'}}
                                       p_{X|V}(i|s), \ i \in \mathcal{X} \\
p_{X|V_{\Delta'}=s_{\Delta'}}^{\inf} &=& \inf_{s \in  \mathcal{S}_{\Delta'}}
                                       p_{X|V}(i|s), \ i \in \mathcal{X}
\end{eqnarray}

$p_{X|V_{\Delta'}=s_{\Delta'}}^{\sup}$ is not, in general, a
probability measure. It is a measure with mass $\geq$ 1 and denotes a
measure which ``dominates'' all probability measures $p_{X|V=s}$ over
the quantization region of $\mathcal{V}$ which contains $s_{\Delta'}$.

$p_{X|V_{\Delta'}=s_{\Delta'}}^{\inf}$ is not, in general, a probability
measure. It is a measure with mass $\leq$ 1. It denotes a measure which 
``is dominated by'' all probability measures $p_{X|V=s}$ over the
quantization region of $\mathcal{V}$ which contains $s_{\Delta'}$.

Intuitively, if we make some continuity assumptions on $p_{X|V=s}$
as $s \in \mathcal{V}$ varies,  then
$p_{X|V_{\Delta'}=s_{\Delta'}}^{\sup}$  and 
$p_{X|V_{\Delta'}=s_{\Delta'}}^{\inf}$
will be close to each other. For small enough $\Delta'$, all $s \in 
\mathcal{S}_{\Delta'}$ are almost the same in the distribution 
induced on $\mathcal{X}$.

Another reason for defining $p_{X|V_{\Delta'}=s_{\Delta'}}^{\sup}$
is that it helps us to do error probability calculations. This is
demonstrated by the following lemma:

\begin{lemma}\label{UpperBound}
Let $p_X$ be a probability distribution on $\mathcal{X}$. Let 
$\mu_X$ be a measure on $\mathcal{X}$ such that 
$\mu_X(i) > p_X(i)$ for all $i \in \mathcal{X}$
(that is, $\mu_X$ dominates $p_X$). Let $q_X$ be another probability 
distribution on $\mathcal{X}$. 

Then, probability that an $n$ length sequence generated iid 
$p_X$ has type $q_X$ 
\begin{equation}
p_X^n(T(q_X)) \leq 2^{-nD(q_X||\mu_X)}
\end{equation}
where $D(q_X||\mu_X)$ is defined in the obvious way,
$D(q_X||\mu_X) \triangleq 
            \sum_{i \in \mathcal{X}} q_X(i) log \frac{q_X(i)}{\mu_X(i)}$
\end{lemma}
Proof: 
\begin{eqnarray*}
& & p_X^n(T(q_X)) \\
&\leq& 2^{-nD(q_X||p_X)} \ \mbox{(by method of types)} \\ 
              &\leq& 2^{-nD(q_X||\mu_X)} \ 
                                \mbox {(trivial by definition of
                                                         $D(q_X||\mu_X)$)}
\end{eqnarray*}

This lemma gives us a way of upper bounding the error probability of a type
class when we do not know the generating distribution, but have an upper
bound on the same, and this is precisely the situation we are in.

We define $p_{X|V_{\Delta'}=s_{\Delta'}}^{\mbox{\tiny{avg}}}$ as the 
probability measure obtained by normalizing 
$p_{X|V_{\Delta'}=s_{\Delta'}}^{\sup}$.

If we have some continuity conditions (which we will make rigorous later)
on $p_{X|V}$, as measures, 
$p_{X|V_{\Delta'}=s_{\Delta'}}^{\sup}$,
$p_{X|V_{\Delta'}=s_{\Delta'}}^{\inf}$,
$p_{X|V_{\Delta'}=s_{\Delta'}}^{\mbox{\tiny{avg}}}$,
$\{ p_{X|V=s}, s \in \mathcal{S}_{\Delta'} \}$ will be quite close to each
other.

Also, the distributions $p_{V_{\Delta'}}$ and 
$p_{X|V_{\Delta'}=s_{\Delta'}}^{\mbox{\tiny{avg}}}$
result in a probability distribution on 
$(V_{\Delta'}, X)$ which we denote
by $p_{V_{\Delta'} X}^{\mbox{\tiny{avg}}}$.

Next, we state the codebook formation and decoding rule:

\textbf{Codebook Construction}: Generate $2^{nR}$ codewords iid $p_{X|V}$. 
This is the codebook $\mathcal{C}$.

\textbf{Decoding}: Fix $\epsilon> 0$. 
Restrict attention to those codewords $x_1^n$ that 
$(x_1^n, v_{1\Delta'}^n)$ have an empirical type $q_{X,V_{\Delta'}}$
that is $p_{V_{\Delta'}X}^{\mbox{\tiny{avg}}}$ typical.
Denote this restricted set of codewords by $\mathcal{C}_R$.

Let $y_1^n$ denote the output of the attacker. If there is a unique
$x_1^n$ in the restricted codebook $\mathcal{C}_R$ that is at an
average distortion less than or equal to $D$ from the output sequence,
declare that $x_1^n$ was transmitted, else declare error.

We impose the following technical condition\footnote{It can be shown
  to be satisfied for any joint distribution for $X,V$ that satisfies
  weak convergence in that $p(X|V=s_n) \to p(X|V=s)$ whenever $s_n \to
s$.} on 
$p_{X|V_{\Delta'}=s_{\Delta'}}^{\sup}$ and 
$p_{X|V_{\Delta'}=s_{\Delta'}}^{\inf}$,
which captures mathematically, the closeness of 
$p_{X|V = s_1}$ and  $p_{X|V = s_2}$ for $s_1$ and $s_2$ close.

\textbf{Technical Condition:} $\forall i \in \mathcal{X}$
\begin{equation}\label{TechnicalConditionFinite}
\lim_{\Delta' \to 0} \max_{s_{\Delta'} \in \mathcal{V}_{\Delta'}}
\left |p_{X|V_{\Delta'}=s_{\Delta'}}^{\sup}(i|s_{\Delta'})
    -  p_{X|V_{\Delta'}=s_{\Delta'}}^{\sup} (i|s_{\Delta'}) \right | 
                                  = 0 
\end{equation}
\vspace{0.1in}
This condition says that 
$p_{X|V_{\Delta'}=s_{\Delta'}}^{\sup}(i|s_{\Delta'})
    -  p_{X|V_{\Delta'}=s_{\Delta'}}^{\sup} (i|s_{\Delta'}) \to 0 \ 
 \mbox{as} \Delta' \to 0$ \textbf{uniformly} over all partitions of 
$\mathcal{V}$.

We now do the probability of error calculations.

It is easy to check that with the above decoding rule, the
probabilities of error event $E_2 \to 0$ as $n \to
\infty$. For $E_1$, all that is required is for $n$ to be large enough
while $\epsilon$ is also large enough relative to $\Delta'$ so that
$[p_{V_{\Delta'}X}^{\inf} - \frac{\epsilon}{2},
p_{V_{\Delta'}X}^{\sup} + \frac{\epsilon}{2}] \in
p_{V_{\Delta'}X}^{\mbox{\tiny{avg}}} \pm \epsilon$. At that point, the
weak law of large numbers is enough to guarantee what is desired. 

For $\Pr(E_3)$,  we follow the steps in the proof of Case 2
($\mathcal{X}$ finite   $\mathcal{Y}$ finite,  $\mathcal{V}$ finite)
(Theorem~\ref{MainTheoremConditional}) and use Lemma~\ref{UpperBound}
to replace $p_{X|V}$ with $p_{X|V_{\Delta'}}^{\sup}$. It follows that
we can transmit at rates 
\begin{equation}
R < 
\lim_{\epsilon \to 0}
 \inf_ {\begin{array}{c} q_{V_{\Delta'}X} \in 
                     p_{V_{\Delta'}X}^{\mbox{\tiny{avg}}} \pm \epsilon \\
    Ed_{q_{V_{\Delta'}XY}}(X,Y) \leq D \end{array}} 
 \hspace{-.7cm}     D \left ( q_{XY|V_{\Delta'}}||p_{X|V_{\Delta'}}^{\sup} 
                         q_{Y|V_{\Delta'}}|q_{V_{\Delta'}} \right )
\end{equation}
First thing that we need to take care of 
$p_{X|V_{\Delta'}}^{\sup}$ appearing above - we want to somehow replace
it by $p_{X|V_{\Delta'}}^{\mbox{\tiny{avg}}}$. Using the
technical condition (\ref{TechnicalConditionFinite}), 
it is easy to see that there is a function $h$ such that we can
transmit $R < $

\begin{equation}
\lim_{\epsilon \to 0}
\hspace{-0.3cm} \inf_ {\begin{array}{c} q_{V_{\Delta'}X} \in 
                     p_{V_{\Delta'}X}^{\mbox{\tiny{avg}}} \pm \epsilon \\
    Ed_{q_{V_{\Delta'}XY}}(X,Y) \leq D \end{array}} 
    \hspace{-.7cm} D \left ( q_{XY|V_{\Delta'}}||p_{X|V_{\Delta'}}^ {\mbox{\tiny{avg}}}
                         q_{Y|V_{\Delta'}}|q_{V_{\Delta'}} \right )
       - h(\Delta')
\end{equation}
where $h(\Delta') \to 0$ as $\Delta' \to 0$.
The first term above is the same as that appearing in the proof of 
Case 2, the conditional case with $\mathcal{X}, \mathcal{Y},
\mathcal{V}$ finite, Equation \ref {ConditionalEquation}. It follows
that we can transmit at all rates 
\begin{equation}
R < 
\inf_ {\begin{array}{c} (V_{\Delta'},X) \sim 
                     p_{V_{\Delta'}X}^{\mbox{\tiny{avg}}} \\
    Ed(X,Y) \leq D \end{array}} 
     \hspace{-1.3cm}                I(X; Y|V_{\Delta'}) - h(\Delta')
   = R_{X|V_{\Delta'}}(D) - h(\Delta')
\end{equation}
Now, $\lim_{\Delta' \to 0} R_{X|V_{\Delta'}}(D) - h(\Delta')
 = R_{X|V}(D)$ (we need to use the technical condition
(\ref{TechnicalConditionFinite}) for proving this), and it follows that
we can transmit at all rates less than $R_{X|V}(D)$.

The sequence in which we choose $n, \epsilon, \Delta'$ depending
on the rate $R$ and the probability of error $p_e$ is
\begin{enumerate}

\item 
Choose $\Delta'$ small enough so that 
$R < R_{X|V_{\Delta'}}(D) - h(\Delta')$

\item Choose $\epsilon$ small enough such that 
$R < \inf_{(V_{\Delta'},X) \in 
p_{V_{\Delta'}X}^{\mbox{\tiny{avg}}} \pm \epsilon}
R_{X|V_{\Delta'}}(D) - h(\Delta')$

\item Choose $n$ large enough so that sum of error probabilities 
of events $E_1, E_2, E_3 < p_e$.

\end{enumerate}

Finally, we consider Case 6, that of $\mathcal{X}, \mathcal{Y},
\mathcal{V}$ non-finite. This is just a mixture of decoding rules for 
Case 4 ($\mathcal{X}$ non-finite, $\mathcal{Y}$ non-finite,
$\mathcal{V}$ finite) and the previous case, Case 5 
( $\mathcal{X}$ finite, $\mathcal{Y}$ finite , $\mathcal{V}$ non-finite).

First quantize $\mathcal{X}, \mathcal{Y}$ to size $\Delta$. This way,
we get $p_{X_{\Delta} | V}$. This reduces the problem to previous case where 
$\mathcal{X}$ and $\mathcal{Y}$ are finite and by combining the
decoding rules of  Case 4 and Case 5, it is easy to see that we can
transmit at all rates $R < R_{X_{\Delta}|V}(D + g(\Delta))$ where
$g(\Delta)$ is defined analogous to that in Case 3. 

Taking $\Delta \to 0$, it follows that we can transmit at all rates
$R < R_{X|V}(D)$.

Clearly, the technical condition in place of
(\ref{TechnicalConditionFinite}) in this case of $\mathcal{X}, \mathcal{Y}$ 
non-finite, but bounded support, is: $\forall x_{\Delta} \in
\mathcal{X_{\Delta}}$ 
\begin{equation}\label{TechnicalConditionInfinite}
\lim_{\Delta' \to 0} \max_{s_{\Delta'} \in \mathcal{V}_{\Delta'}}
\left |p_{X_{\Delta}|V_{\Delta'}=s_{\Delta'}}^{\sup}(x_{\Delta}|s_{\Delta'})
    -  p_{X_{\Delta}|V_{\Delta'}=s_{\Delta'}}^{\sup} 
                                           (x_{\Delta}|s_{\Delta'}) \right | 
                          = 0 
\end{equation}
This is just saying that the technical condition of the finite 
$\mathcal{X}$ case should hold for all partitions of 
$\mathcal{X}$ in this non-finite case.

The sequence in which we choose $\epsilon, n, \Delta, \Delta'$ 
to achieve a rate R and probability of error $< p_e$ is
\begin{enumerate}
\item Choose $\Delta$ small enough so that $R < R_{X_{\Delta}|V}(D +
  g(\Delta))$. 

\item Choose $\Delta'$ small enough so that 
      $R < R_{X_{\Delta}|V_{\Delta'}}(D + g(\Delta)) - h(\Delta')$.

\item Choose $\epsilon$ small enough so that 
$R < \inf_{(V_{\Delta'},X_{\Delta}) \in 
p_{V_{\Delta'}X_{\Delta}}^{\mbox{\tiny{avg}}} \pm \epsilon}
R_{X_{\Delta}|V_{\Delta'}}(D + g(\Delta)) - h(\Delta')$

\item Choose $n$ large enough so that sum of error probabilities
  caused by events $E_1, E_2, E_3$ add up to less than $p_e$.
\end{enumerate}
 
Next, we state (without proof) sufficient conditions for the technical
conditions, Equations (\ref{TechnicalConditionFinite}) and
(\ref{TechnicalConditionInfinite}) to hold.

\begin{enumerate}
\item Case 5, that is, $\mathcal{X}, \mathcal{Y}$ finite, 
$\mathcal{V}$ non-finite: The following weak convergence condition is 
sufficient for the technical condition (\ref{TechnicalConditionFinite})
to hold:
\begin{equation}
s_{\alpha} \to s \implies p_{X|V=s_{\alpha}} \stackrel{w}
{\longrightarrow}p_{X|V=s}
\end{equation}

\item Case 6, that is, $\mathcal{X}, \mathcal{Y}, 
\mathcal{V}$ non-finite: what we want is that after discretizing 
$\mathcal{X}$ and $\mathcal{Y}$, the same technical condition should hold.
Assuming that $p_{X|V=s}$ \textbf{have densities}, the above condition,
\begin{equation}
s_{\alpha} \to s \implies p_{X|V=s_{\alpha}} \stackrel{w}
{\longrightarrow}p_{X|V=s}
\end{equation}
is sufficient for the technical condition (\ref{TechnicalConditionInfinite})
to hold. 
\end{enumerate}

\subsection{Unbounded support} \label{sec:unbounded}

The compact support condition is what allowed us to use quantization
to reduce everything to the finite-alphabet case where the method of
types could work since the number of possible types grew only
polynomially in the block-length $n$. Dealing with this requires
an appropriate truncation argument. For space reasons, we merely
sketch the essential ideas here:

\begin{enumerate}
 \item Pick a small $\delta > 0$.
 \item Pick a sufficiently large compact region ${\mathcal X}_c \times
   {\mathcal V}_c$ (with the obvious modifications if there is no
   coverstory) so that it satisfies the following properties:
  \begin{itemize} 
  \item $P({\mathcal X}_c \times {\mathcal V}_c) \geq 1 - \delta$
  \item $P({\mathcal X}_c | V = s) \geq 1 - \delta$ for all $s \in
    {\mathcal V}_c$ 
  \item Let $X_c,V_c$ be the random variables $X,V$ conditioned on
    their values lying within the compact region ${\mathcal X}_c \times
   {\mathcal V}_c$. Then $R_{X_c|V_c}(D) \geq (1-\delta)
   R_{X|V}(D)$. 
  \end{itemize}
  Given this, the distribution for $P(X|V=s)$ can be written as a
  convex combination $(1-\delta)P_{X_c|V_c = s} + \delta P'_{X|V_c =
  s}$ for some other distribution $P'_{X|V_c = s}$. 
 \item Employ a two-part strategy for generating the random
   codebook. First, we classify positions in the codebook as ``clean''
   or ``dirty'' or ``bad'':
   \begin{itemize}
    \item Mark as ``dirty'' all positions where $V_t$ is not in
          ${\mathcal V}_c$. 
    \item Flip a commonly random iid biased coin with $\delta$
          probability of coming up heads for each position. Mark as
          ``bad'' all positions where the coin turns up heads.
    \item All remaining positions are ``clean.''
   \end{itemize}
   Next, we generate the $2^{nR}$ random codewords iid using
   $P_{X_c|V_c}$ in the clean positions. For dirty positions, we draw
   from $P_{X|V}$ while bad positions are drawn from $P'_{X|V_c}$. The
   resulting codewords look as though they are drawn from $P_{X|V}$.

 \item For decoding, look at only the clean positions. If their number
   is less than $(1-4\delta)n$, declare error. Beyond that, we
   treat it as in the previous cases dealing with compact support,
   using the appropriate quantization and nearest typical neighbor
   decoding.
\end{enumerate}

In terms of the probability of error, there is now a new error event
$E_0$ which corresponds to there being more than $4\delta n$ bad or
dirty positions. By the weak law of large numbers (since bad and dirty
positions arrive no faster than a Bernoulli processes with expected
rate $2\delta$), this cannot happen very often and so $P(E_0) \to 0$
as $n \to \infty$. 

The other terms in the probability of error can be bounded by
pretending that the attacker knows not only the dirty positions, but
also the bad ones. Assume it also knows that our decoding rule is
going to ignore all the dirty and bad positions. With this knowledge,
the worst thing it can do is choose to allocate no distortion to those
positions and spend that distortion over the clean positions. However,
this only increases average distortion by a factor
$\frac{1+4\delta}{1-4\delta}$ over the clean positions that figure in
the decoding process. By choosing $\delta$ sufficiently small, we can
be sure that $R < R(D(\frac{1+4\delta}{1-4\delta}))$.  Everything else
proceeds as before.

\section{Stationary-Ergodic Sources} \label{WithMemory} So far, the
information-embedding arguments seemed to depend strongly on the
assumption of memorylessness. This is what allowed the method-of-types
to be used.  To deal with more general sources with memory, we can
just apply a trick similar to the truncation argument in
Section~\ref{sec:unbounded}. Once again, in the interest of space, we
simply sketch the key ideas in the context of finite-alphabet
rate-distortion problems.

Suppose that the source process $\{X_t\}$ is
stationary\footnote{Since time for us starts at $1$, assume that it
  has been initialized into its stationary distribution.} and
ergodic. In such cases, the rate-distortion and conditional
rate-distortion functions are defined in terms of limits of longer and
longer finite-horizon problems $X_1^t$. So, for any $t$
sufficiently long, then $R < R_{X}(D)$ implies also that $tR <
R_{X_1^t}(tD)$. But before we simply pick a $t$ long enough, we need
to impose a technical condition that requires the process to ``mix''
appropriately uniformly fast towards its stationary distribution.

Assume that for every $\lambda > 0, \beta > 0$, there exists a uniform
delay $\tau$ so that for all $t>0$, all possible values\footnote{All
  the arguments here immediately generalize to the conditional
  rate-distortion case if the technical condition holds uniformly over
  all possible realizations for the cover-story sequence $V_1^\infty$.
  Essentially, we want to capture the idea that the cover-story should
  not be able to force the $\{X_t\}$ process to strongly remember what
  it did in its distant past. This condition can be relaxed so that it
  is only required to hold for most realizations of the cover-story 
  process.} $x_1^t$, all $k > 0$, and all measurable subsets $A$ of
${\mathcal X}^k$:
\begin{equation} \label{eqn:fadingmemorycondition}
 P(X_{t+d}^{t+d+k-1} \in A | X_1^t = x_1^t) =
 (1-\lambda)P_{\mbox{stat}}^\beta(X_1^k \in A) + \lambda P'(A)
\end{equation}
where $P'$ is a probability measure that can depend explicitly on
$t,d,x_1^t$ while $P_{\mbox{stat}}^\beta$ is a measure that does
not have any such dependence and is within $\pm \beta$ of the
stationary probability distribution for the original process. 

Essentially, (\ref{eqn:fadingmemorycondition}) just captures the idea
that the process has fading memory and that if we wait long enough,
the process will return to its stationary distribution regardless of
what values the process might have taken in the past. It is easy to
verify that (\ref{eqn:fadingmemorycondition}) holds for all
finite-state stationary ergodic Markov chains\footnote{Because they
  must mix exponentially fast based on the second largest eigenvalue
  of the transition matrix.} as well as hidden Markov models with
an underlying finite-state stationary ergodic Markov chain. 

With this condition, the codebook construction proceeds in the
following sequence:
\begin{enumerate}
 \item Pick small enough $\lambda, \beta$
 \item Based on the technical condition, calculate the required delay
   $d$ to make the process ``forget'' its past.
 \item Pick a $t$ sufficiently long so that $\frac{t}{t+d}$ is close
   to $1$, and the finite horizon rate-distortion function is close to
   its infinite-horizon limit. 
 \item Segment time regularly with $t$ time units of
   potentially embedded data followed by $d$ time units of dead-time.
 \item Use common-randomness to generate Bernoulli$(\lambda)$ random
   variables used to mark $t$-long slots as being bad. This is done
   for the entire codebook, not on a codeword by codeword basis.
 \item For the codewords, independently generate the $t$-long slots
   that are not bad by drawing from the stationary distribution for
   $X_1^t$. Draw bad slots using $P'$ from
   (\ref{eqn:fadingmemorycondition}) and the prefix of the
   codeword\footnote{If the block code is intended to be used over and
     over again, then in general the dead-times must be interpolated
     in a way that takes into account what was transmitted in the
     distant past. This is not a problem for Markov or Hidden Markov
     processes.} so far.
 \item Generate the $d$-length dead-time slots in between by sampling
   from the appropriate conditional distribution once the following
   $t$-long slot has been chosen. 
 \end{enumerate}

It is clear that every codeword is thus a simulation of the original
process with memory. Conditioned on knowing where the good slots of
length $t$ are, the process is iid from both the encoder and decoder's
point of view and so reverts to the previous case. The decoder can
focus entirely on the good slots viewed as an iid process. Once again,
the probability of having fewer than a $(1-2\lambda)$ proportion of
good slots goes to zero. Decoding error can be bounded by supposing
that the attacker knew which slots were good and what time-segments
were ``dead-time.'' Thus, the attacker can choose to concentrate all
its distortion on the good slots. This increases the average
distortion by a factor of at most
$\frac{t+d}{t}(\frac{1+2\lambda}{1-2\lambda})$ --- which is as close
to $1$ as we want.  

\bibliographystyle{IEEEtran}
\bibliography{IEEEabrv,./MyMainBibliography} 
\end{document}